\documentclass{aipproc}

\usepackage{natbib,graphicx}
\layoutstyle{6x9}

\begin{document}

\title{SN\,2002lt and GRB\,021211: a SN/GRB Connection at $z = 1$}

\author{M. Della Valle}{address={INAF, Osservatorio Astrofisico di
  Arcetri, largo E. Fermi 5, 50125 Firenze, Italy}}
\author{D. Malesani}{address={International School for Advanced Studies
  (SISSA/ISAS), via Beirut 2-4, 34014 Trieste, Italy}}
\author{S. Benetti}{address={INAF, Osservatorio Astronomico di Padova,
  vicolo dell'Osservatorio 5, 35122 Padova, Italy}}
\author{V. Testa}{address={INAF, Osservatorio Astronomico di Roma, via
  Frascati 33, 00040 Monteporzio (Roma), Italy}}
\author{M. Hamuy}{address={Carnegie Observatories, 813 Santa Barbara
  Street, Pasadena, California 91101, USA}}
\author{L.A. Antonelli}{address={INAF, Osservatorio Astronomico di Roma,
  via Frascati 33, 00040 Monteporzio (Roma), Italy}}
\author{G. Chincarini}{address={INAF, Osservatorio Astronomico di Brera,
  via E. Bianchi 46, 23807 Merate (Lc), Italy},
  altaddress={University of Milano--Bicocca, Department of Physics,
  Piazza delle Scienze, 20126 Milano, Italy}}
\author{G. Cocozza}{address={INAF, Osservatorio Astronomico di Roma, via
  Frascati 33, 00040 Monteporzio (Roma), Italy},
  altaddress={University of Roma Tor Vergata, Dept. of Physics, via
  d. Ricerca Scientifica, 00133 Roma, Italy.}}
\author{S. Covino}{address={INAF, Osservatorio Astronomico di Brera,
  via E. Bianchi 46, 23807 Merate (Lc), Italy}}
\author{P. D'Avanzo}{address={INAF, Osservatorio Astronomico di Brera,
  via E. Bianchi 46, 23807 Merate (Lc), Italy}}
\author{D. Fugazza}{address={INAF, Osservatorio Astronomico di Brera,
  via E. Bianchi 46, 23807 Merate (Lc), Italy},
  altaddress={INAF, TNG, Roque de Los Muchachos, PO box 565, 38700 Santa
  Cruz de La Palma, Spain}}
\author{G. Ghisellini}{address={INAF, Osservatorio Astronomico di Brera,
  via E. Bianchi 46, 23807 Merate (Lc), Italy}}
\author{R. Gilmozzi}{address={ESO, Alonso de Cordova 3107, Casilla
  19001, Vitacura, Santiago, Chile}}
\author{D. Lazzati}{address={Institute of Astrophysics, University of
  Cambridge, Madingley Road, CB3 0HA Cambridge, UK}}
\author{E. Mason}{address={ESO, Alonso de Cordova 3107, Casilla 19001,
  Vitacura, Santiago, Chile}}
\author{P. Mazzali}{address={INAF, Osservatorio Astronomico di Trieste,
  via Tiepolo 11, 34131 Trieste, Italy}}
\author{L. Stella}{address={INAF, Osservatorio Astronomico di Roma, via
  Frascati 33, 00040 Monteporzio (Roma), Italy}}

\begin{abstract}
  We present spectroscopic and photometric observations of the afterglow
  of GRB\,021211 and the discovery of its associated supernova,
  SN\,2002lt. The spectrum shows a broad feature (FWHM$\mbox{}=
  150$~\AA), around 3770~\AA{} (in the rest-frame of the GRB), which we
  interpret as Ca H+K blueshifted by 14\,400 km/s. Potential sources of
  contamination due to the host galaxy and/or residuals of telluric
  absorption have been analyzed and ruled out. Overall, the spectrum
  shows a suggestive resemblance with the one of the prototypical
  type-Ic SN\,1994I. This might indicate that GRBs are produced also by
  standard type-Ic supernov\ae.
\end{abstract}

\maketitle

\section{Introduction.}

After long years of study, we have now some convincing evidence that
long-duration gamma-ray bursts (GRBs) are produced by the death of a
massive star. The earliest hint was the spatial and temporal coincidence
between SN\,1998bw and GRB\,980425 (at $z = 0.0085$;
\cite{Galama98}). However, this association was hardly representative of
the whole class of GRBs: GRB\,980425 was indeed a very dim event, its
gamma-ray energy being lower than that of classical GRBs by $\sim 4$
orders of magnitude. SN\,1998bw was also a peculiar event, belonging to
the class of the so-called `hypernov\ae': it showed unusual expansion
velocities and a very high luminosity, the latter being the effect of
copious Nickel production (see e.g. \cite{Nomoto00}).

Very recently, however, SN\,2003dh was discovered associated with
GRB\,030329 (\cite{Stanek03,Hjorth03,Matheson03}). An intensive spectral
monitoring showed a strict similarity between SN\,2003dh and SN\,1998bw,
thereby conclusively proving that hypernov\ae{} can generate classical
GRBs. SN\,2003dh was however somewhat fainter than SN\,1998bw
(\cite{Mazzali03}).

We present here photometric and spectroscopic observations
highlightening the association between GRB\,021211 and SN\,2002lt
(\cite{DV03}). GRB\,021211 (\cite{Crew03}) was a rather dim event, its
total (isotropic) gamma-ray energy being $E_\mathrm{iso} = (6 \pm 0.5)
\times 10^{51}$~erg, at the low end of the energy distribution of GRBs
(\cite{Frail01}). Prompt optical observations allowed an early discovery
of the optical afterglow, just few minutes after the GRB onset
(e.g. \cite{Li03,Fox03}). The afterglow also turned out to be quite
faint, about 3 magnitudes in the $R$ optical band. The redshift was
determined to be $z = 1.006$ (\cite{Vreeswijk02}). Optical/NIR colors
showed little or no extinction local to the host galaxy.  The intrinsic
faintness of the afterglow made this event a good candidate for
searching a supernova component.

\section{Data and analysis}

\begin{figure}
  \includegraphics[width=0.4\columnwidth]{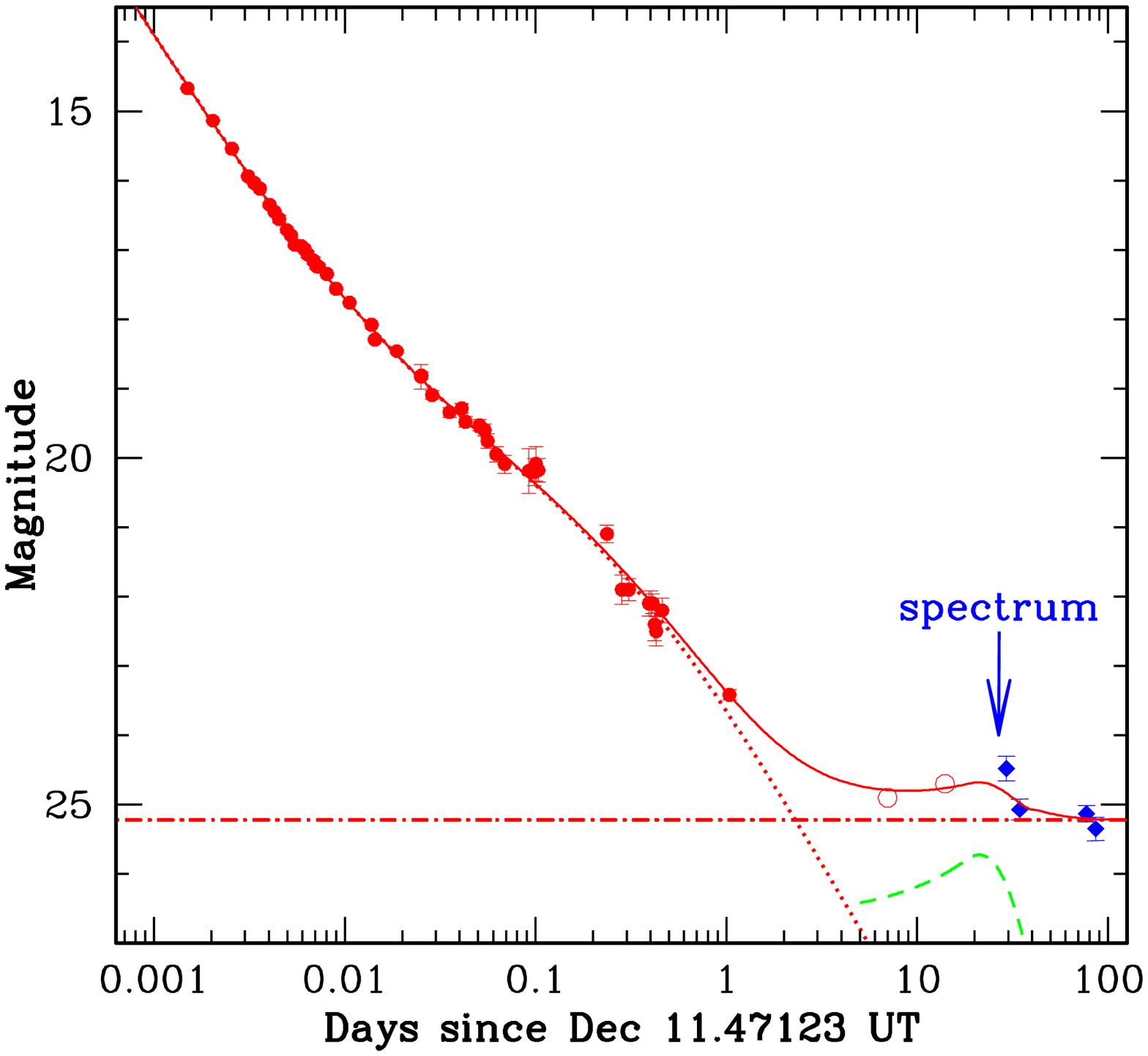}\hspace{1cm}
  \includegraphics[width=0.4\columnwidth]{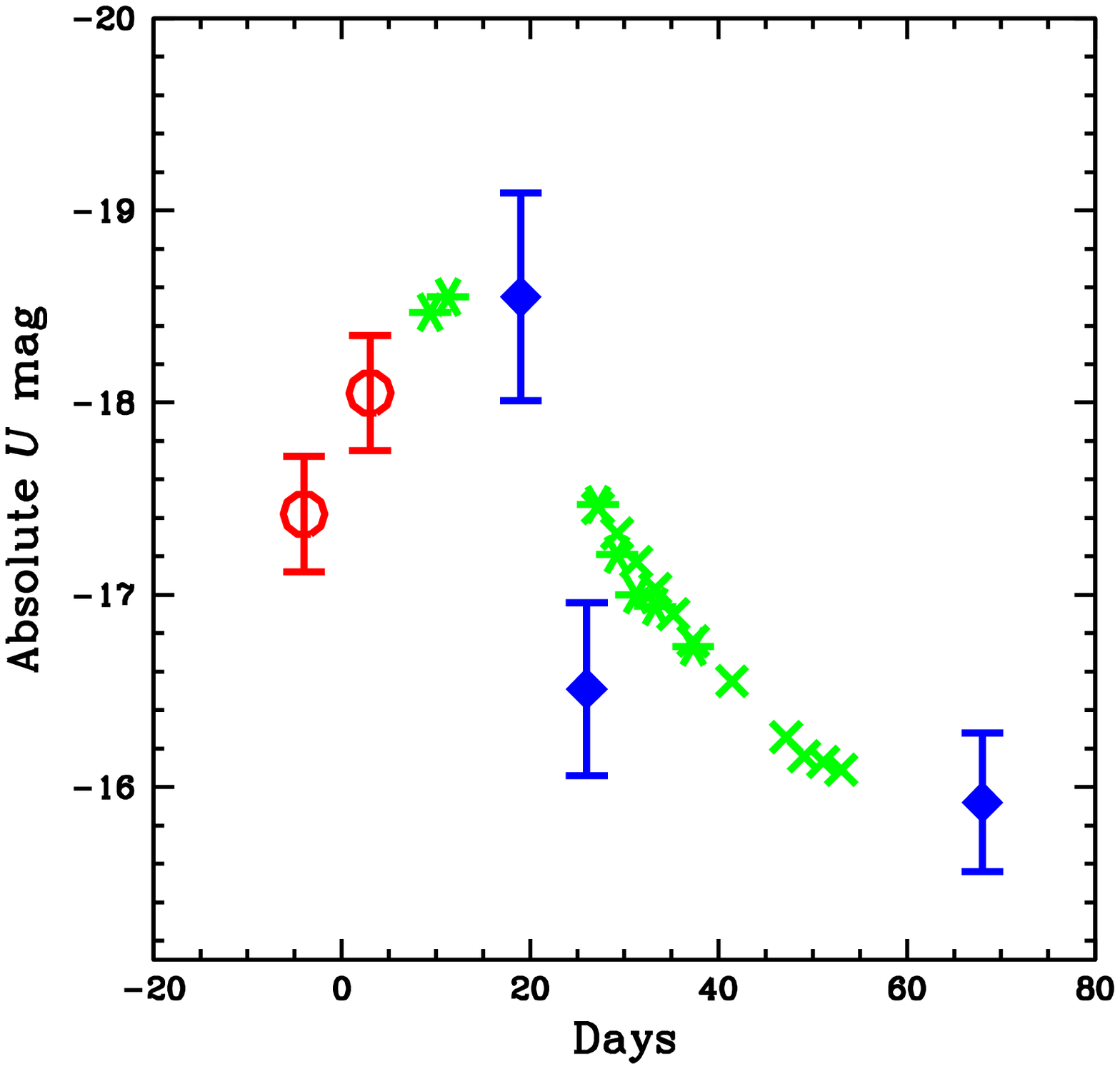}
  \caption{{\bf Left panel}. $R$-band light-curve of the optical
  afterglow of GRB\,021211. Data come from our observations (diamonds),
  the literature (filled circles; \cite{Li03,Fox03,Pandey03}), and HST
  measurements (open circles; \cite{Fruchter02}). The dotted and
  dot-dashed lines represent the afterglow and host contribution
  respectively. The dashed line shows the light curve of SN\,1994I
  reported at $z = 1.006$ and dereddened with $A_V = 2$ (from
  \cite{Lee95}). The solid line shows the sum of the three
  contributions. {\bf Right panel}. Comparison of the $U$-band light
  curves of the rebrightening of GRB\,021211, after subtracting the host
  galaxy contribution (filled diamonds and open circles) and of
  SN\,1994I, dereddened with $A_V = 1.8$ (asterisks; crosses have been
  obtained from the $V$-band lightcurve after assuming
  $U-V=\mbox{}$constant). Errorbars represent 3-$\sigma$ errors.}
\end{figure}

We observed the optical afterglow of GRB\,021211 with the ESO VLT--UT4
equipped with the FORS\,2 instrument, during the period January -- March
2003 (20 -- 100 days after the GRB). Low-resolution spectroscopy was
performed on Jan~8.27 UT.

Photometric data show a rebrightnening of the afterglow starting $\sim
15$~days after the burst (\cite{Fruchter02}) and reaching the maximum,
$R \sim 24.5$, during the first week of January. The contribution of the
host galaxy, estimated from our late-epoch images, is $R = 25.22 \pm
0.10$. Therefore, the intrinsic magnitude of the bump was $R = 25.24 \pm
0.38$. The significance of the rebrightening is at the 4-$\sigma$
level. At the time of the maximum, the afterglow contribution,
extrapolated from earlier epochs, is smaller than $5\%$ under the most
conservative assumptions. This suggests that the bump is powered by some
different component.

\begin{figure}
  \includegraphics[width=0.8\columnwidth]{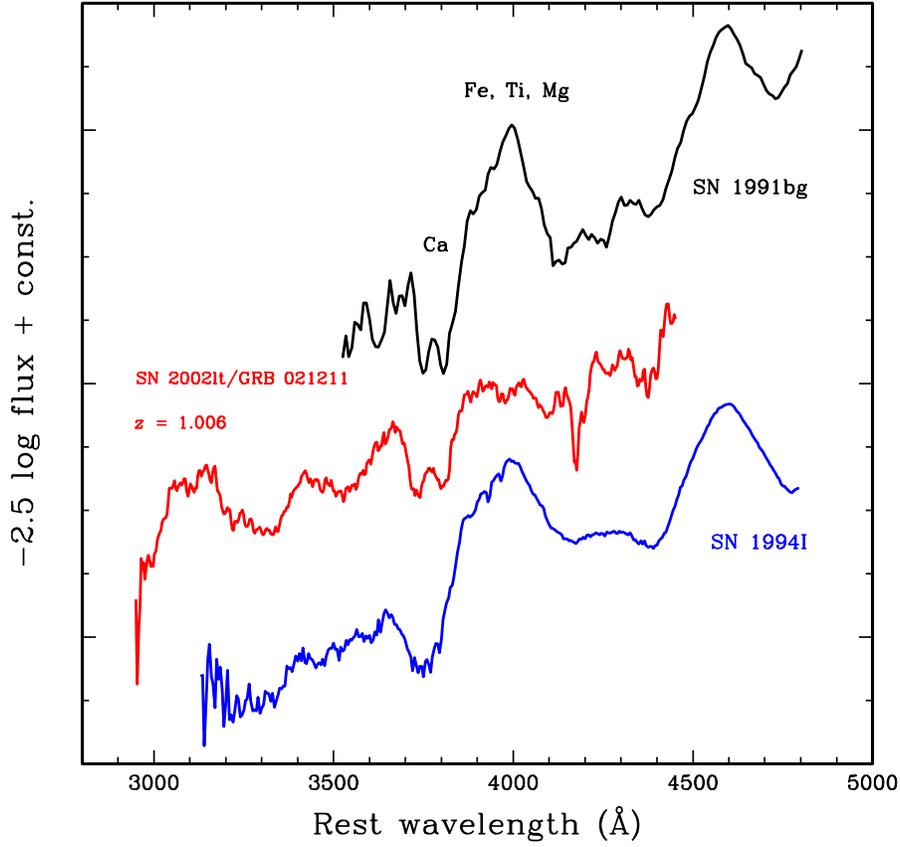}
  \caption{Spectrum of the afterglow+host galaxy of GRB\,021211 (middle
  line), taken on Jan~8.27~UT (27~days after the burst). For comparison,
  the spectra of SN\,1994I (type Ic, bottom) and SN\,1994I (type Ia,
  top) are displayed, both showing the Ca absorption.\label{fg:spec}}
\end{figure}

To investigate the nature of the rebrightening, we obtained a spectrum
with VLT+FORS\,2. The reduced spectrum covered the range of wavelengths
$(6000 - 9000)$~\AA{} at an acceptable S/N ($> 3$). The resolution was
about 19~\AA, and the integration time was $4 \times 1$~h with a seeing
of $0''.6$ -- $1''.4$. We confirm the detection of the emission line at
$\lambda = 7472.9$~\AA{} (\cite{Vreeswijk02}), which may be interpreted
as [O\,II] 3727~\AA{} in the rest frame, leading to a determination of
the redshift $z = 1.006$.

\begin{figure}\centering
  \includegraphics[width=0.6\columnwidth]{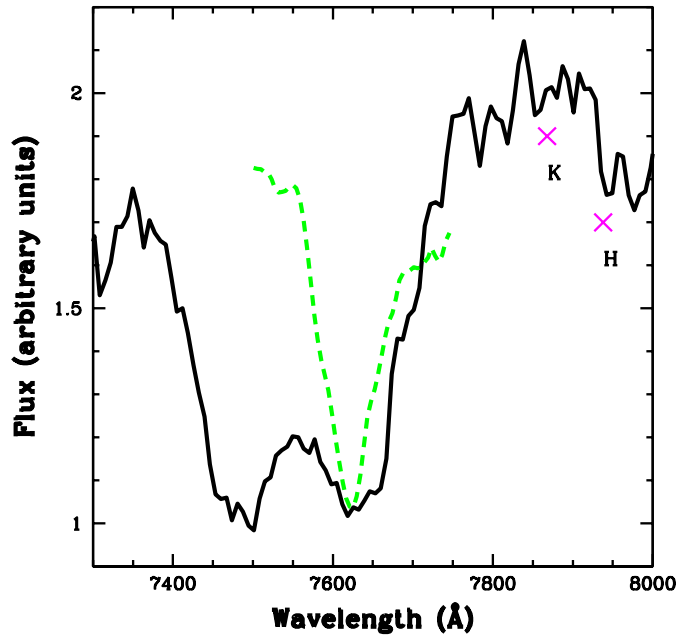}
  \caption{Comparison of the absorption deep (solid line) in GRB\,021211
  and the telluric feature at $\lambda \sim 7600$~\AA{} (dashed line),
  smoothed with the same boxcar filter 55~\AA{} wide. The two patterns
  do not match, thereby confirming that the feature is intrinsic to the
  afterglow.\label{fg:telluric}}
\end{figure}

Fig.~\ref{fg:spec} shows our spectrum smoothed and cleaned from the
emission line [O\,II]. The spectrum of the afterglow is characterized by
broad low-amplitude undulations blueward and redward of a broad
absorption, the minimum of which is measured at $\sim 3770$~\AA{} (in
the rest frame of the GRB), whereas its blue wing extends up to $\sim
3650$~\AA. The absorption feature in our spectrum is a characteristic
signature of the SN ejecta (see SN\,1991bg and SN\,1994I in
Fig.~\ref{fg:spec}) and it is due to Ca\,II H+K absorption lines. The
blueshifts corresponding to the minimum of the absorption and to the
edge of the blue wing imply velocities of $v \sim 14\,400$~km/s and $v
\sim 23\,000$~km/s respectively. The more convincing resemblance is
found with SN\,1994I, a prototypical type-Ic event, 9~days after its
$B$-band maximum (\cite{Filippenko95}).  It is difficult to explain such
broad absorption feature in terms of other components. It was argued
that the telluric absorption at $\sim 7600$~\AA{} can contaminate, by
chance, the absorption feature. Fig.~\ref{fg:telluric} shows that even
in the case that the subtraction of the telluric line was not effective
(which is not the case), its profile cannot reproduce the broad and
double-structured deep in the spectrum. Also in Fig.~\ref{fg:telluric}
the position of the rest-frame Ca H+K absorption lines is marked
(crosses), showing that no contamination comes from the host galaxy.

\section{Implications}

The spectroscopic features observed during the rebrightening of the
afterglow of GRB\,021211 and the similarity of its lightcurve with the
one of SN\,1994I indicate that the bump was indeed powered by a
supernova. This is therefore the third GRB (second in chronological
order) for which a SN association was spectroscopically confirmed. The
IAU dubbed this event SN\,2002lt (\cite{DV03b}). SN\,2001ke and
GRB\,011121 are another possible case of SN/GRB association, although no
`typical' SN features have been detected in the spectrum
(\cite{Garnavich03}).

\begin{figure}\centering
  \includegraphics[width=0.8\columnwidth]{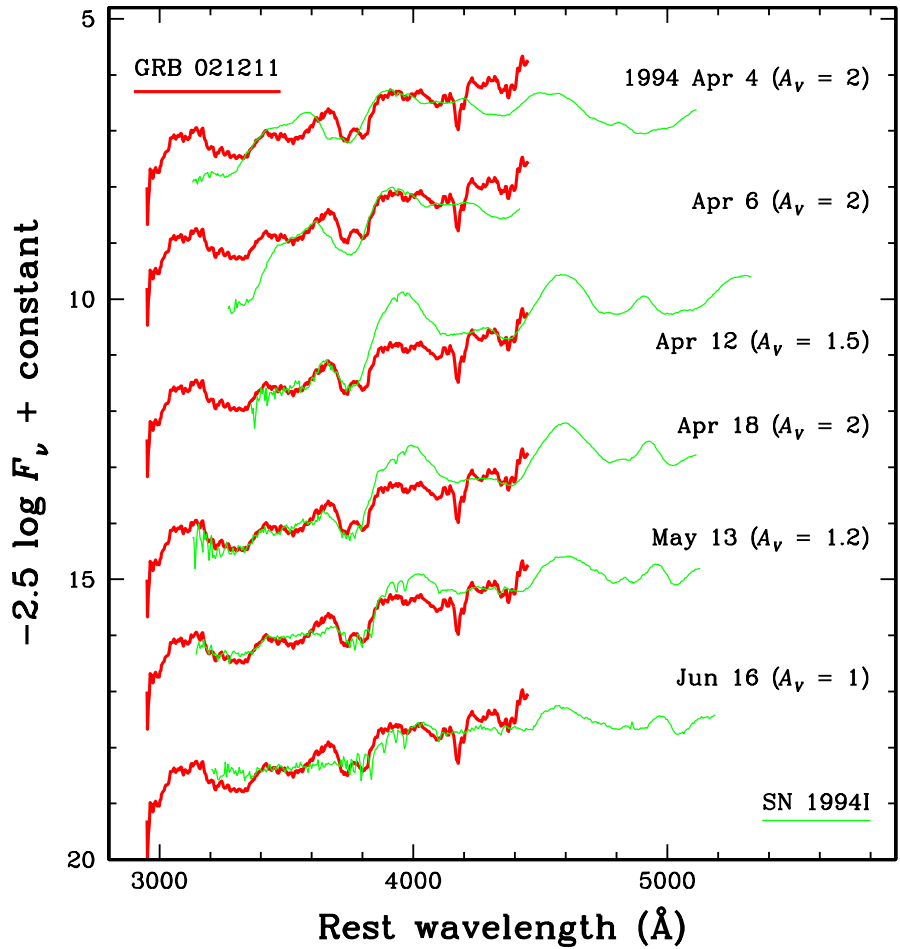}
  \caption{Comparison of our spectrum with the ones of SN\,1994I at
  various epochs (spectra taken from \cite{Filippenko95}). For
  reference, SN\,1994I reached its $B$-band maximum on
  1994~Apr~9.\label{fg:timespec}}
\end{figure}

{\bf Supernova dating}. Using SN\,1994I as a template, our photometric
and spectroscopic data allow us to estimate the time at which the SN
exploded, and to compare it with the GRB onset time. Due to the limited
wavelength coverage and to the lack of multi-time spectroscopic
observations, our spectrum yields only a shallow contrain, suggesting
that the SN went off between $\sim\mbox{}$50 and 0 days before the GRB
(see Fig.~\ref{fg:timespec}).

Information from the photometry yields more stringent limits. After
assuming SN\,1994I as a template, the best match is achieved if the SN
and the GRB exploded simultaneously or separated, at most, by a few
days. We stress however that also in this case the dataset is not rich
enough to set firm bounds. Moreover, this result also depends upon the
assumed rising time of the template SN, which was quite short
($\sim\mbox{}$12 days) for SN\,1994I (\cite{Iwamoto94}).

{\bf GRB progenitor}. It is also interesting to note that SN\,1994I, the
spectrum of which provides the best match to that observed in
GRB\,021211, is a typical type-Ic event rather than an exceptional
1998bw-like object, as the one proposed for association with GRB\,980425
and GRB\,030329 (\cite{Galama98,Stanek03,Hjorth03}). If the SN
associated with GRB\,021211 indeed shared the properties of SN\,1994I,
this would open the interesting possibility that GRBs may be associated
with standard type-Ic SNe, and not only with the more powerful events
known as `hypernov\ae'.  However, we should note that the recently
studied SN\,2002ap (\cite{Mazzali02}) showed significantly broader lines
than our case and this difference vanished after maximum, such that it
may be not easy to distinguish between the two types of SNe.

\end{document}